\documentclass[journal]{IEEEtran}
\usepackage{ifpdf}
\usepackage{cite}
\usepackage{amsmath}
\usepackage{algorithmic}
\usepackage{array}
\usepackage{fixltx2e}
\usepackage{stfloats}
\usepackage{bm}
\usepackage{url}
\usepackage[pdftex]{graphicx}
\usepackage{xcolor}

\ifCLASSINFOpdf
\else
\fi
\begin{document}

\title{Ultra-low power on-chip learning of speech commands with phase-change memories}
%
%
% author names and IEEE memberships
% note positions of commas and nonbreaking spaces ( ~ ) LaTeX will not break
% a structure at a ~ so this keeps an author's name from being broken across
% two lines.
% use \thanks{} to gain access to the first footnote area
% a separate \thanks must be used for each paragraph as LaTeX2e's \thanks
% was not built to handle multiple paragraphs
%

\author{Venkata~Pavan~Kumar~Miriyala, Masatoshi~Ishii
        % <-this % stops a space

\thanks{This work has been submitted to the IEEE for possible publication.
Copyright may be transferred without notice, after which this version may
no longer be accessible.}
\thanks{The authors are with IBM Research Tokyo, Shin-Kawasaki, Japan, 212-0032.}
\thanks{\emph{V. P. K. Miriyala and M. Ishii contributed equally to this work}. \emph{(Corresponding author: M. Ishii, email: ishiim@jp.ibm.com)}}}
\maketitle

% As a general rule, do not put math, special symbols or citations
% in the abstract or keywords.
\begin{abstract}
Embedding artificial intelligence at the edge (edge-AI) is an elegant solution to tackle the power and latency issues in the rapidly expanding Internet of Things. As edge devices typically spend most of their time in sleep mode and only wake-up infrequently to collect and process sensor data, non-volatile in-memory computing (NVIMC) is a promising approach to design the next generation of edge-AI devices. Recently, we proposed an NVIMC-based neuromorphic accelerator using the phase change memories (PCMs), which we call as Raven. In this work, we demonstrate the ultra-low-power on-chip training and inference of speech commands using Raven. We showed that Raven can be trained on-chip with power consumption as low as 30~$\mu$W, which is suitable for edge applications. Furthermore, we showed that at iso-accuracies, Raven needs 70.36$\times$ and 269.23$\times$ less number of computations to be performed than a deep neural network (DNN) during inference and training, respectively. Owing to such low power and computational requirements, Raven provides a promising pathway towards ultra-low-power training and inference at the edge.
\end{abstract}

% Note that keywords are not normally used for peerreview papers.

% For peer review papers, you can put extra information on the cover
% page as needed:
% \ifCLASSOPTIONpeerreview
% \begin{center} \bfseries EDICS Category: 3-BBND \end{center}
% \fi
%
% For peerreview papers, this IEEEtran command inserts a page break and
% creates the second title. It will be ignored for other modes.
\IEEEpeerreviewmaketitle

\section{Introduction}
\label{sec:sec_I}

%\IEEEPARstart{I}{n}
In this era of rapidly expanding internet of things (IoT), embedding artificial intelligence (AI) at the edge (edge-AI) is an elegant solution to tackle the cost, bandwidth, power, latency, and privacy issues arising from edge-to-cloud computing \cite{SDENG2020, Zhang2019, Lee2018, YDAI2019}. At present, deep neural networks (DNNs) provide the best classification accuracies in solving many AI problems such as image classification, pattern/object recognition, speech recognition, etc. \cite{geirhos2017, Seifert2017, FENG2019}. As a result, DNNs are commonly used to embed AI at the edge. Usually, the training of DNNs is performed in the cloud, the learned weights are transferred to the edge, and only inference is performed at the edge. The reason is that the training of DNNs typically requires backpropagation of end results throughout the network \cite{rumelhart1986}, and it needs large amounts of memory and computational resources. However, IoT environments such as autonomous driving, security surveillance, and smart cities continuously change over time. If training for such environments is performed in the cloud, a large amount of data needs to be transmitted to the cloud, which leads to higher costs, increased latencies, and lower bandwidths \cite{bali2013, Hu2018}. Alternatively, performing training at the edge can be a promising approach to achieve continuous real-time learning with reduced cost, latency, and bandwidth concerns.

Recently, spiking neural networks (SNNs) have emerged as potential computing paradigms for enabling AI at the edge \cite{koo2020, reiter2020neuromorphic}. Inspired by the information processing mechanisms in the brain, the data in SNNs is encoded and processed in the form of binary spikes. As processing time increases, the spiking activity in SNNs reduces drastically \cite{khoei2020, allred2020explicitly}. Moreover, SNNs are event-driven, which means computations are performed only when the neurons emit/receive the spikes \cite{Blouw, PATI20319}. Furthermore, SNNs can be trained using the spike-time-dependent plasticity (STDP) learning rule \cite{annurevneuro2008}. STDP is a localized learning rule, where the weights are updated based on the relative timing of spikes emitted by a neuron and of those that it receives. Therefore, due to the STDP-based localized learning ability, sparse spiking activity, and event-driven computations, the SNNs facilitate ultra-low-power training and inference of data at the edge.

On the other hand, one major concern in today's edge devices \cite{Scass2019, XXu2017, 1Scass2020} is that they are designed based on the conventional von Neumann architecture with separate memory and processing units. As a result, the data must be transferred between memory and processing units to perform any operation. Such data movement results in long inference delays and additional power overheads. In addition, there exists a significant gap between memory and processor speeds. The widely used main memories-dynamic random-access memories (DRAMs) \cite{Feldmann2020} are several orders lower than their processing counterparts. As a result, the overall performance of the system is limited more by the slow memories rather than processors.

One solution that has recently emerged is in-memory computing (IMC), where some computational tasks are performed within the memory subsystem \cite{Verma2019, Zhange2019, Xsi2020, joel2020}. When provided with inputs, the data in IMC engines (IMCEs) can be updated and processed \emph{in-situ} by eliminating the latency and power consumed to transfer data between memory and processing units in conventional von Neumann architectures. Presently, many existing and emerging memory technologies can be used to design the IMCEs \cite{Verma2019, Zhange2019, Xsi2020, joel2020}. Several works have recently demonstrated IMCEs based on DRAMs and static random-access memories (SRAMs) \cite{Zhange2019, Xsi2020, joel2020, seshadri2017, XXin2020}. Though SRAMs and DRAMs facilitate relatively fast read, write and compute operations, they are volatile memories (i.e. the memory subsystem must be always ON for data to be retained). As edge devices typically spend most of their time in sleep mode, the use of volatile memories results in significant standby power consumption. 

In contrast, the non-volatile in-memory computing (NVIMC) is a crucial design technique for enabling ultra-low power edge devices with reduced latencies \cite{chen2019edge, wang2020, Gao2020, asebastian2020, nandakumar2020, stefano2018, iboybat2018}. The data in NVIMC engines (NVIMCEs) is retained even if the power is turned off. Thus, the NVIMCEs can be powered down to achieve near-zero standby power consumption when the device is in deep sleep mode. If the device wakes up, data in the NVIMCE can be updated and processed \emph{in-situ}. The non-volatile memory (NVM)-based crossbar array is a promising design technique to accelerate the neural networks with massive parallelism \cite{asebastian2020, mhu2016, HLue2019,PDu2019,Yxiang}. Recently, we proposed a non-volatile phase change memory (PCM)-based crossbar architecture \cite{Ishii2019} for accelerating the SNNs in memory. For convenience, we call this architecture as Raven for the rest of this paper. 

In this work, using the devices, circuits, and architectures of Raven (i.e. are proposed in \cite{Ishii2019}), we demonstrate ultra-low-power on-chip training and inference of speech commands. First, we considered the Google’s speech commands dataset \cite{webarticle5} and converted the audio files into the Mel-frequency cepstral coefficient (MFCC) images \cite{Mukh2020}. To learn and classify these images, we accelerated the spiking restricted Boltzmann machines (RBMs) with event-driven contrastive divergence (CD) based learning \cite{Emre2014} on Raven (i.e. using software simulations). Our simulation results show that Raven can be trained on-chip with power consumption as low as 30~$\mu$W, which is suitable for edge applications. Next, we also compared the classification accuracies of our work with the results obtained from DNNs \cite{warden2018speech, li2019speech, patil2019}, commonly used for speech command recognition.

The rest of this paper is structured as follows. Section~\ref{sec:sec_II} introduces the devices, circuits, and architectures of Raven (i.e. are proposed in \cite{Ishii2019}). Section~\ref{sec:sec_III} presents the design strategies implemented to achieve on-chip training and inference of speech commands using Raven. Section~\ref{sec:sec_IV} introduces the PCM hardware-aware spiking RBM simulator used to demonstrate the speech command recognition using Raven. Section~\ref{sec:sec_V} presents the results and discussion on speech command recognition. Finally, Section~\ref{sec:sec_VI} concludes this chapter.
% The very first letter is a 2 line initial drop letter followed
% by the rest of the first word in caps.
% 
% form to use if the first word consists of a single letter:
% \IEEEPARstart{A}{demo} file is ....
% 
% form to use if you need the single drop letter followed by
% normal text (unknown if ever used by the IEEE):
% \IEEEPARstart{A}{}demo file is ....
% 
% Some journals put the first two words in caps:
% \IEEEPARstart{T}{his demo} file is ....
% 
% Here we have the typical use of a "T" for an initial drop letter
% and "HIS" in caps to complete the first word.

\section{Phase-change memory based synaptic circuits and arrays}
\label{sec:sec_II}

PCMs typically exist in either an amorphous phase or in one of the several crystalline phases \cite{gwburr2010, ANUPAM2020, hpwong2010}. When the PCM is in the amorphous phase, a high resistance-state is sensed. When the PCM is in one of the crystalline phases, a low resistance-state is sensed. Furthermore, the phase/resistance of PCMs can be modified electrically based on the joule heating mechanism. Recently, a PCM cell having 200-1000 states, extremely low resistance drift coefficient, and highly linear changes in the conductance is demonstrated experimentally \cite{wkim2019}. Therefore, owing to the linearity, non-volatility, a large number of resistance states, and high-yield manufacturability, PCMs have recently been the subject of great interest for different applications such as embedded memory \cite{kim2020, arnaud2018, webarticle9}, in-memory processing \cite{asebastian2020,stefano2018, Gallo2020, Wang_2019}, neuromorphic computing \cite{Ishii2019,wright2011,gwburr2017, iboybat2018}, etc. 
\begin{figure}[h]
\centering
\includegraphics[width=3.35in]{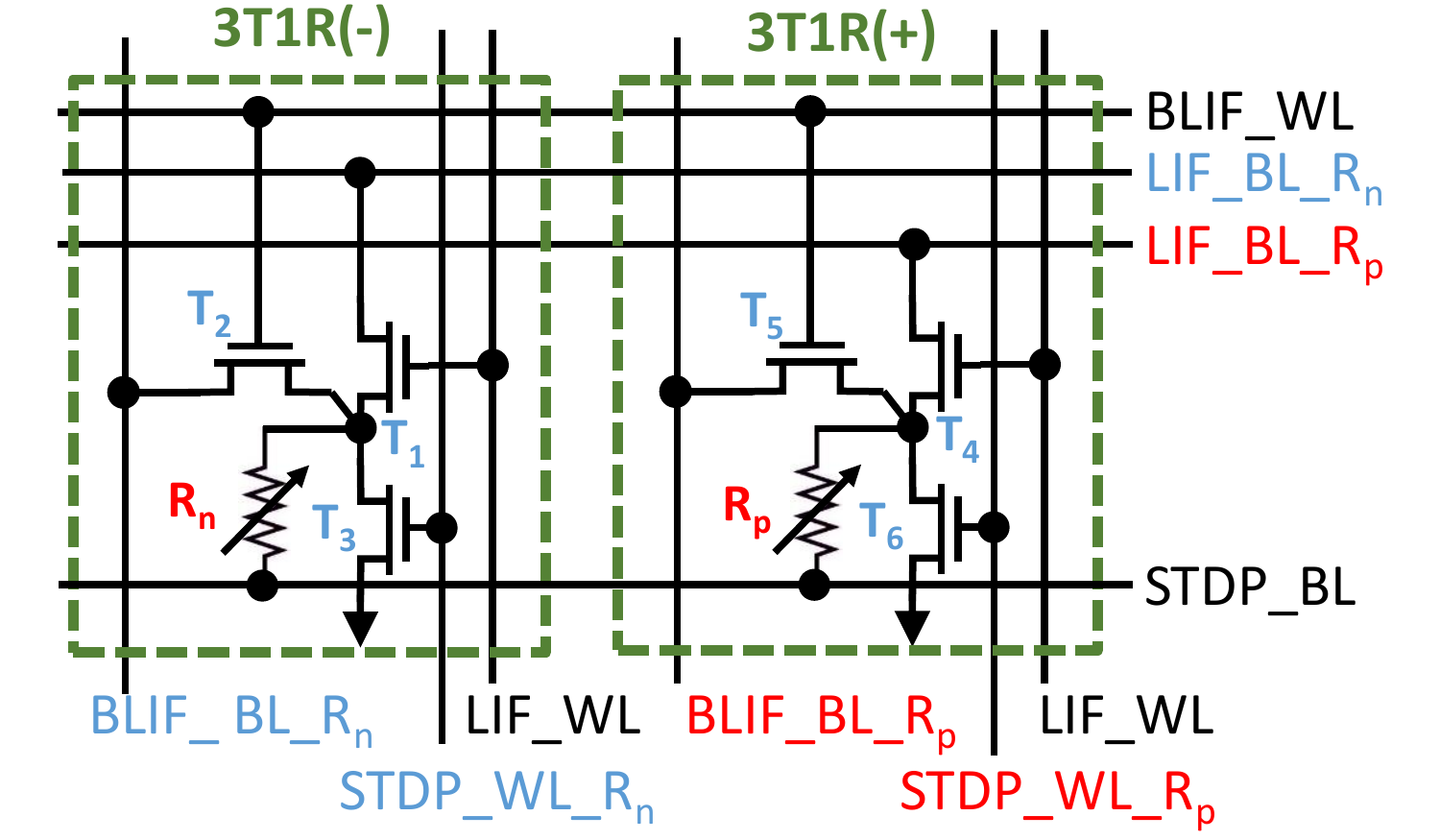}
\caption{Circuit equivalent of the PCM synapse comprising of two 3T1R circuits \cite{Ishii2019}. The two variable resistors, $R_{p}$ and $R_{n}$ are designed using the non-volatile PCMs. The synaptic weight is stored as the difference of analog conductance between $R_{p}$ and $R_{n}$. To access the weight electrically, two currents are passed through the $R_{p}$ and $R_{n}$ from current mirror circuit. The difference of the flowing current are sensed in the current mirror circuit by charging and discharging a capacitor in neurons. The voltage drop/gain on the capacitor indicates the magnitude and sign of the synaptic weight. Note that synaptic weight should be defined as conductance ($\it{G}$) rather than resistance ($\it{R}$). However, for convenience, we define the synaptic weight in terms of $\it{R}$ in this work.}
\label{fig:Fig_1}
\addtocounter{figure}{0}
\end{figure}

Recently, we proposed a novel PCM-based synapse comprising of two 3T1R (3 transistors, 1 resistor) circuits \cite{Ishii2019} (See Fig.~\ref{fig:Fig_1}). The two non-volatile PCM-based variable resistors, $R_{p}$ and $R_{n}$, are used to store the signed weight of the synapse. To access the stored weight, two currents are passed through the $R_{p}$ and $R_{n}$ in 3T1R(+) and 3T1R(-) circuits, respectively. The difference between the resistance values of $R_{p}$ and $R_{n}$ determines the magnitude and sign of the weight. In addition, when placed in a neural circuit with pre and postsynaptic spiking neurons, the two 3T1R circuits can enable asynchronous operation of three fundamental mechanisms in SNNs: a) spike propagation from pre to the postsynaptic neuron, b) spike propagation from post to the presynaptic neuron, and c) weight update based on STDP. 
\begin{figure*}[h]
\centering
\includegraphics[width=6.5in]{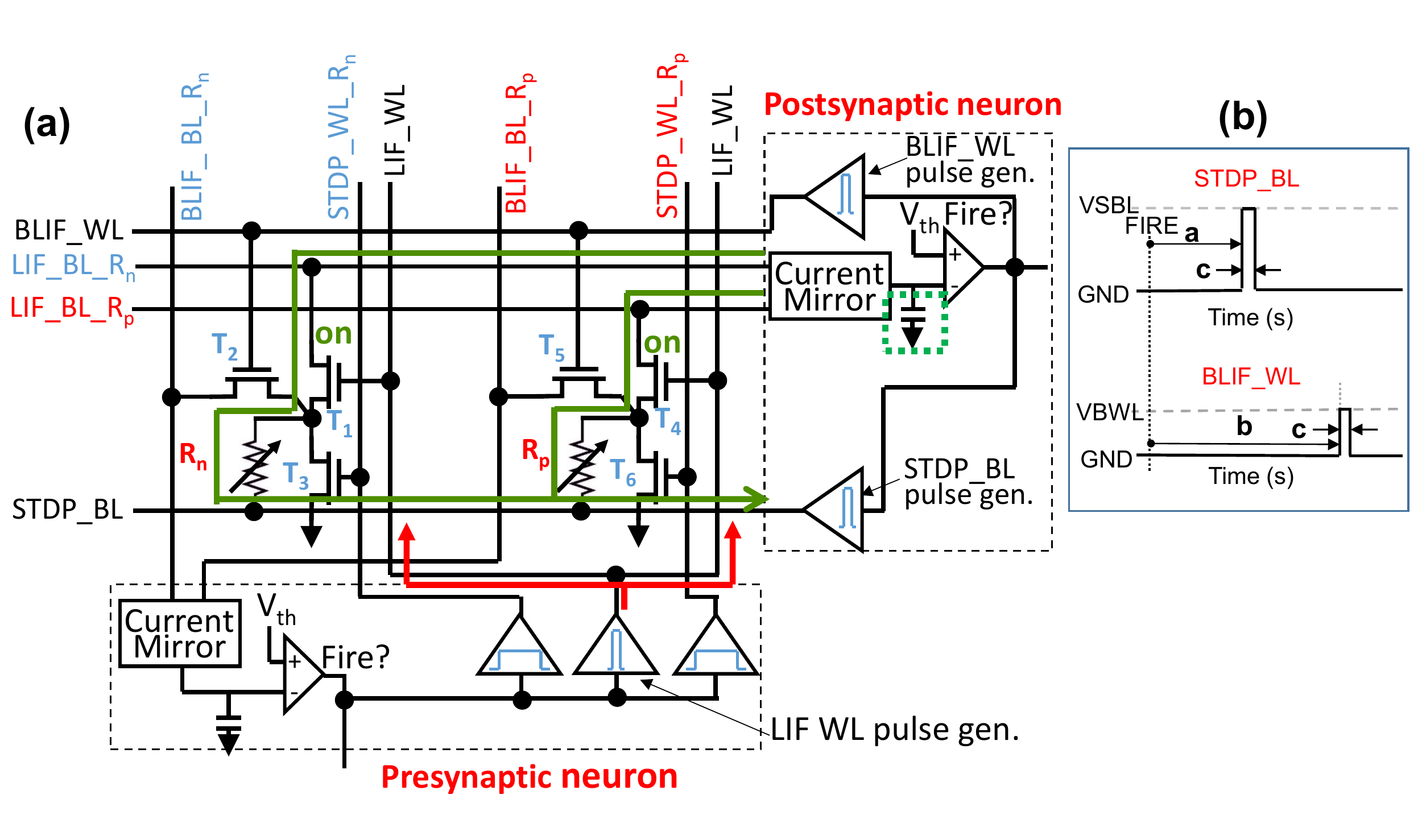}
\caption{Circuit operation of the PCM-based synapse in the presence of pre and postsynaptic neurons \cite{Ishii2019}. (a) when spikes propagate from pre to the postsynaptic neuron. If the potential of postsynaptic neuron exceeds the threshold, spikes with predefined pulse timing (b) will be generated.}
\label{fig:Fig_2_a}
\addtocounter{figure}{0}
\end{figure*}

As mentioned in Section~\ref{sec:sec_I}, the capacitor-based LIF neurons \cite{Ishii2019} are used as pre/postsynaptic neurons in this work. As shown in Fig.~\ref{fig:Fig_2_a}, the neuron circuits consist of capacitors, current mirrors, comparators, and single-shot pulse generators. The voltage stored in the capacitor ($V_{cap}$) is treated as a membrane potential of the neuron. Using $V_{cap}$, the current mirror circuits charge and discharge the current based on the resistance values of $R_{p}$ and $R_{n}$. Consequently, $V_{cap}$ will be updated. The current mirror circuit configuration can be found in \cite{Ishii2019}. If $V_{cap}$ exceeds the pre-defined threshold voltage, $V_{th}$, the comparator will generate a spike using the subsequent single-shot pulse generator. Additionally, several other circuits are needed and used in this work to implement the refractory, leaky, and reset behaviors of conventional LIF neurons. The additional circuitry is omitted from the Figs.~\ref{fig:Fig_2_a}-\ref{fig:Fig_2_c} for simplicity. The complete circuit configuration can be found in \cite{Ishii2019}. 

Now, let us discuss the three operations of the synapse. First, if the presynaptic neuron fires, a spike will be propagated into the word line, $LIF\_WL$. The red-colored line in Fig.~\ref{fig:Fig_2_a}~(a) highlights the direction of spike propagation. Consequently, the transistors, $T_{1}$, and $T_{4}$ will be ON and, the current flows from current mirror circuit through $R_{p}$ and $R_{n}$ as positive and negative current, respectively. The current directions are highlighted using the green-colored lines in Fig.~\ref{fig:Fig_2_a}~(a) and the amount of current is determined by Ohm's law. Then, the current mirror circuit which is connecting to $LIF\_BL\_R_{p}$ and $LIF\_BL\_R_{n}$ senses the difference of the positive and the negative current by charging and discharging one capacitor in the postsynaptic neuron. By using this differential sensing scheme, $V_{cap}$ is increased or decreased depending on the polarity and value of the synaptic weight with every incoming spike. 

As discussed earlier, if $V_{cap}$ exceeds $V_{th}$, the postsynaptic neuron will fire spikes into lines, $STDP\_BL$ and $BLIF\_WL$. The pulse timing of spikes emitted by the postsynaptic neuron is depicted in Fig.~\ref{fig:Fig_2_a}~(b). First, a spike will be fired into the bit-line, $STDP\_BL$ and this will be used for modifying the resistance values of $R_{p}$ and $R_{n}$. After some delay, the second spike will be fired into the word-line, $BLIF\_WL$. Spikes in $BLIF\_WL$ will be used for transmitting the spiking information from post to the presynaptic neuron. 
\begin{figure*}[hbtp]
\centering
\includegraphics[width=6.5in]{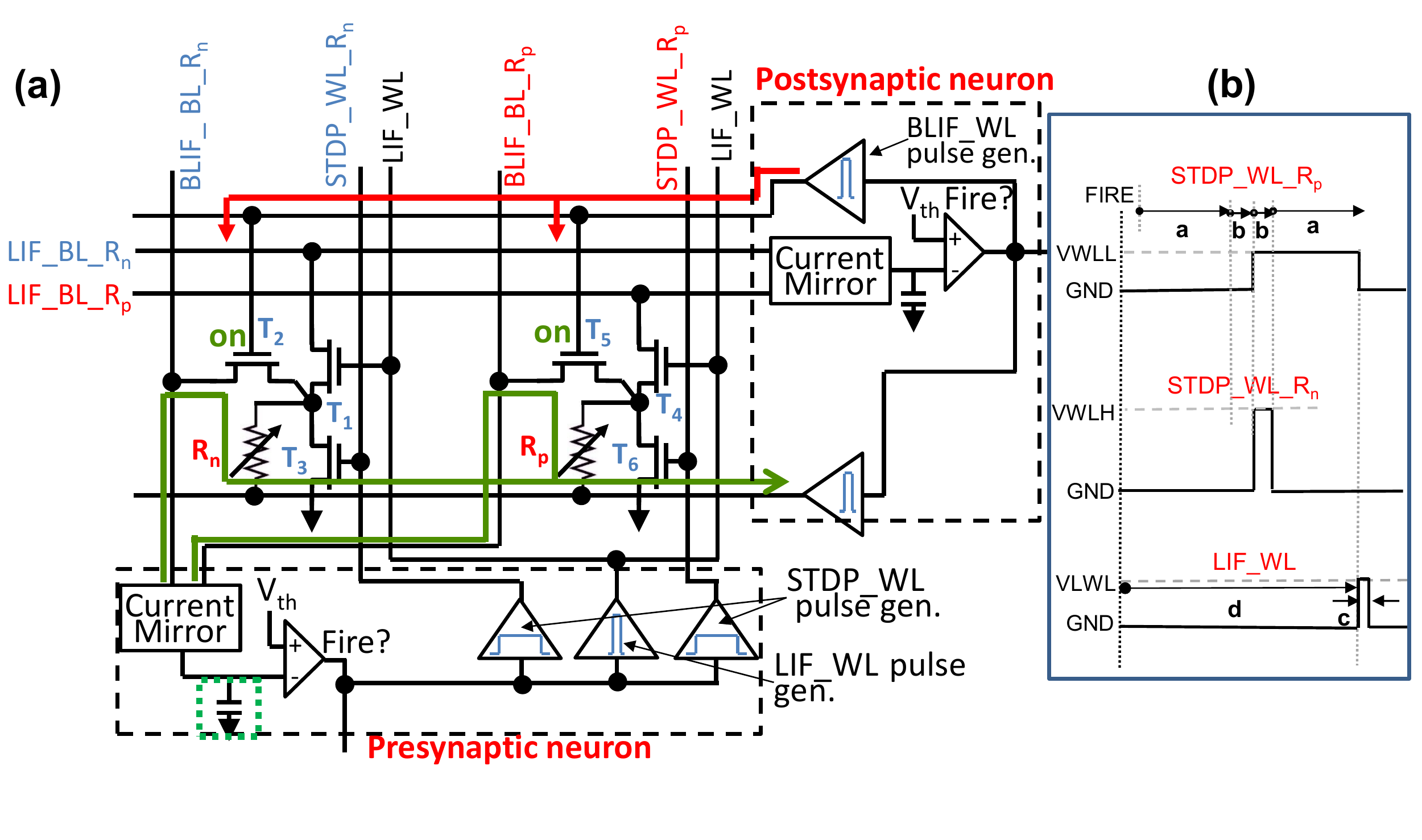}
\caption{Circuit operation of PCM-based synapse in the presence of pre and postsynaptic neurons \cite{Ishii2019}. (a) when spikes propagate from post to the presynaptic neuron. If the potential of presynaptic neuron exceeds the threshold, spikes with predefined pulse timing (b) will be fired.}
\label{fig:Fig_2_b}
\addtocounter{figure}{0}
\end{figure*}
\begin{figure*}[hbtp!]
\centering
\includegraphics[width=6.5in]{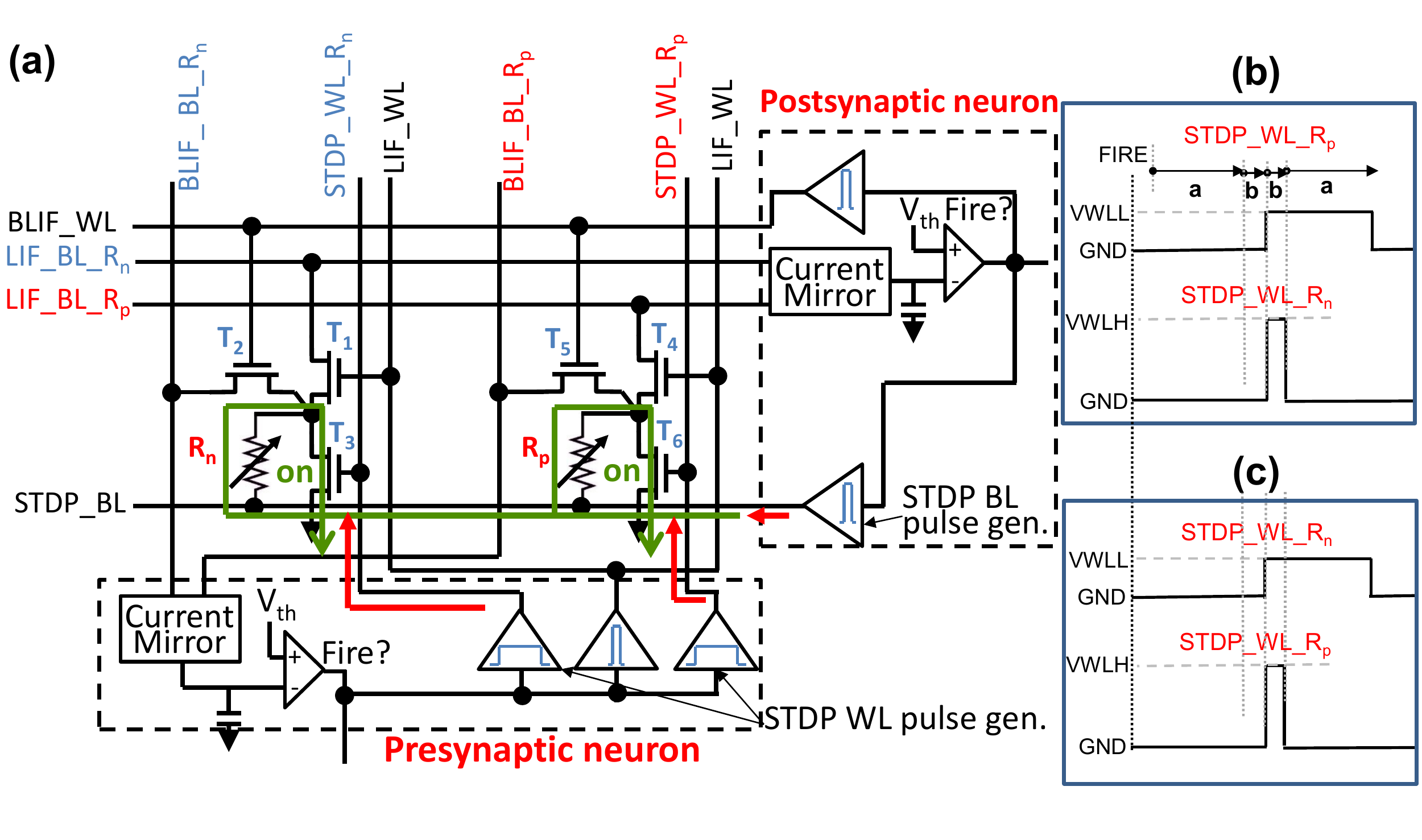}
\caption{(a) the circuit operation during the STDP-based weight update, and the timing diagrams of spikes needed to (b) increase or (c) decrease the weight \cite{Ishii2019}.}
\label{fig:Fig_2_c}
\addtocounter{figure}{0}
\end{figure*}

If a postsynaptic neuron fires, a spike will be propagated into $BLIF\_WL$ (as discussed in the last paragraph). The direction of spike propagation is highlighted by the red-colored line in Fig.~\ref{fig:Fig_2_b}~(a). Consequently, the transistors, $T_{2}$ and $T_{5}$ will be turned ON. Then, the current flows from the current mirror circuit in presynaptic neurons into $STDP\_BL$ through $R_{p}$ and $R_{n}$ as positive and negative current, respectively. Specifically, positive current flows from $BLIF\_BL\_R_{p}$ to $T_{5}$ to $R_{p}$ to $STDP\_BL$ and negative current flows from $BLIF\_BL\_R_{n}$ to $T_{2}$ to $R_{n}$ to $STDP\_BL$. The current directions are highlighted using the green-colored lines (See Fig.~\ref{fig:Fig_2_b}~(a)). The $V_{cap}$ of the presynaptic neuron will be either increased or decreased depending on the resistance values of $R_{p}$ and $R_{n}$. If $V_{cap}$ exceeds the $V_{th}$, the presynaptic neuron fires spikes into lines, $LIF\_WL$, $STDP\_WL\_R_{p}$ and $STDP\_WL\_R_{n}$. The timing of spikes emitted by the presynaptic neuron is depicted in Fig.~\ref{fig:Fig_2_b}~(b). First, the spikes will be fired into $STDP\_WL\_R_{p}$ and $STDP\_WL\_R_{n}$. These spikes will be used for modifying the weight based on the STDP rule. Next, a spike will be fired into $LIF\_WL$, and that can be used for transmitting the spiking information from pre to the postsynaptic neuron.

On the other hand, the spikes propagating through the bit-line, $STDP\_BL$, and word-lines, $STDP\_WL\_R_{p}$, and $STDP\_WL\_R_{n}$ enable modification of the weight. For instance, spikes propagating through $STDP\_WL\_R_{p}$ and $STDP\_WL\_R_{n}$ will turn ON the transistors, $T_{3}$ and $T_{6}$. The directions of spikes propagating in the circuit are highlighted by the red-colored lines (See Fig.~\ref{fig:Fig_2_c}~(a)). Concurrently, if a spike propagates through $STDP\_BL$, new current paths will emerge: a) $STDP\_BL$ to $R_{n}$ to $T_{3}$ to GND and b) $STDP\_BL$ to $R_{p}$ to $T_{6}$ to GND (as highlighted by the green-colored lines in Fig.~\ref{fig:Fig_2_c}~(a)). Depending on the magnitude and duration of currents passing through these paths, the resistances of $R_{p}$ and $R_{n}$ will be modified. For example, when a spike propagating through $STDP\_WL\_R_{p}$ has low magnitude and large pulse width as shown in Fig.~\ref{fig:Fig_2_c}~(b), the resistance value of $R_{p}$ decreases. In other words, the PCM is being set to the crystalline phase (i.e. low resistance state). However, if a spike propagating in $STDP\_WL\_R_{p}$ has high magnitude as shown in Fig.~\ref{fig:Fig_2_c}~(c), the resistance value of $R_{p}$ will increase. In other words, the PCM cell is changed to an amorphous phase (i.e. high resistance state). If spikes propagating through $STDP\_WL\_R_{p}$ and $STDP\_WL\_R_{n}$ as designed with timing diagrams as shown in Fig.~\ref{fig:Fig_2_c}~(b), $R_{p}$ will increase and $R_{n}$ will decrease and the overall weight will be increased. If spikes propagating through $STDP\_WL\_R_{p}$ and $STDP\_WL\_R_{n}$ as designed with timing diagrams as shown in Fig.~\ref{fig:Fig_2_c}~(c), $R_{p}$ will decrease and $R_{n}$ will increase and the overall weight will be decreased.
\begin{figure*}[hbtp!]
\centering
\includegraphics[width=6.5in]{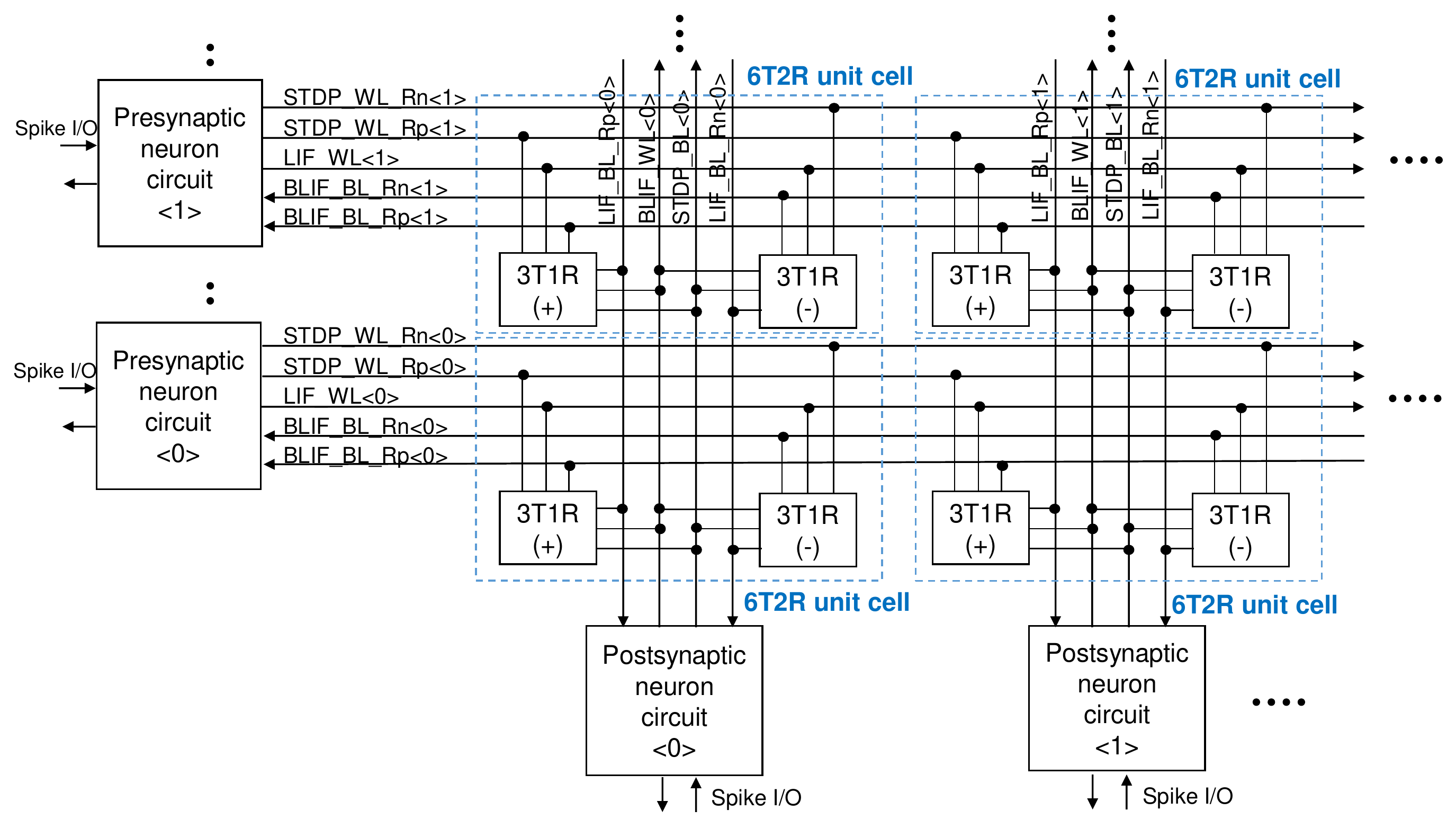}
\caption{The schematic of PCM-based synaptic array connected presynaptic neurons in the left and postsynaptic neurons at the bottom \cite{Ishii2019}, which we call as Raven.}
\label{fig:Fig_3}
\addtocounter{figure}{0}
\end{figure*}

Furthermore, Fig.~\ref{fig:Fig_3} shows the architecture of Raven designed using the above-discussed synaptic and neural circuits.  As shown in Fig.~\ref{fig:Fig_3}, the synaptic circuits are arranged in a crossbar array-like structure with presynaptic neurons in the left and postsynaptic neurons at the bottom. Moreover, the area of an 832$\times$832 array connected to 832 presynaptic neurons in the left and 832 postsynaptic neurons at the bottom is estimated to be 2.20 mm$\times$2.55 mm \cite{Ishii2019}.

\section{Speech Command Recognition using Spiking RBMs}
\label{sec:sec_III}

The Raven circuits and architectures introduced in Section~\ref{sec:sec_II} can be used to demonstrate the on-chip training and inference of speech commands. We will now discuss the design strategies, algorithms, and neural networks used for such a demonstration.    

First, Google’s speech commands dataset is considered \cite{webarticle5} in this work. This dataset contains more than 0.1 million utterances of 30 different words. Importantly, it contains words that can be used as commands in the IoT/robotics applications, \emph{e.g.} stop, go, left, right, up, down, on, off, yes, no, etc. Besides, it also contains the recordings of spoken digits from 0 to 9, various kinds of background noise, and few random utterances (e.g. ``happy'', ``bird'', ``horse'', ``tree'', ``wow'', etc). Each audio file in the dataset is one-second-long and is sampled at 16 kHz. Throughout this work, 500 audio files of each command are used to create the training datasets and 250 different audio files of each command are used for creating the test datasets.
\begin{figure*}[h]
\centering
\includegraphics[width=6.5in]{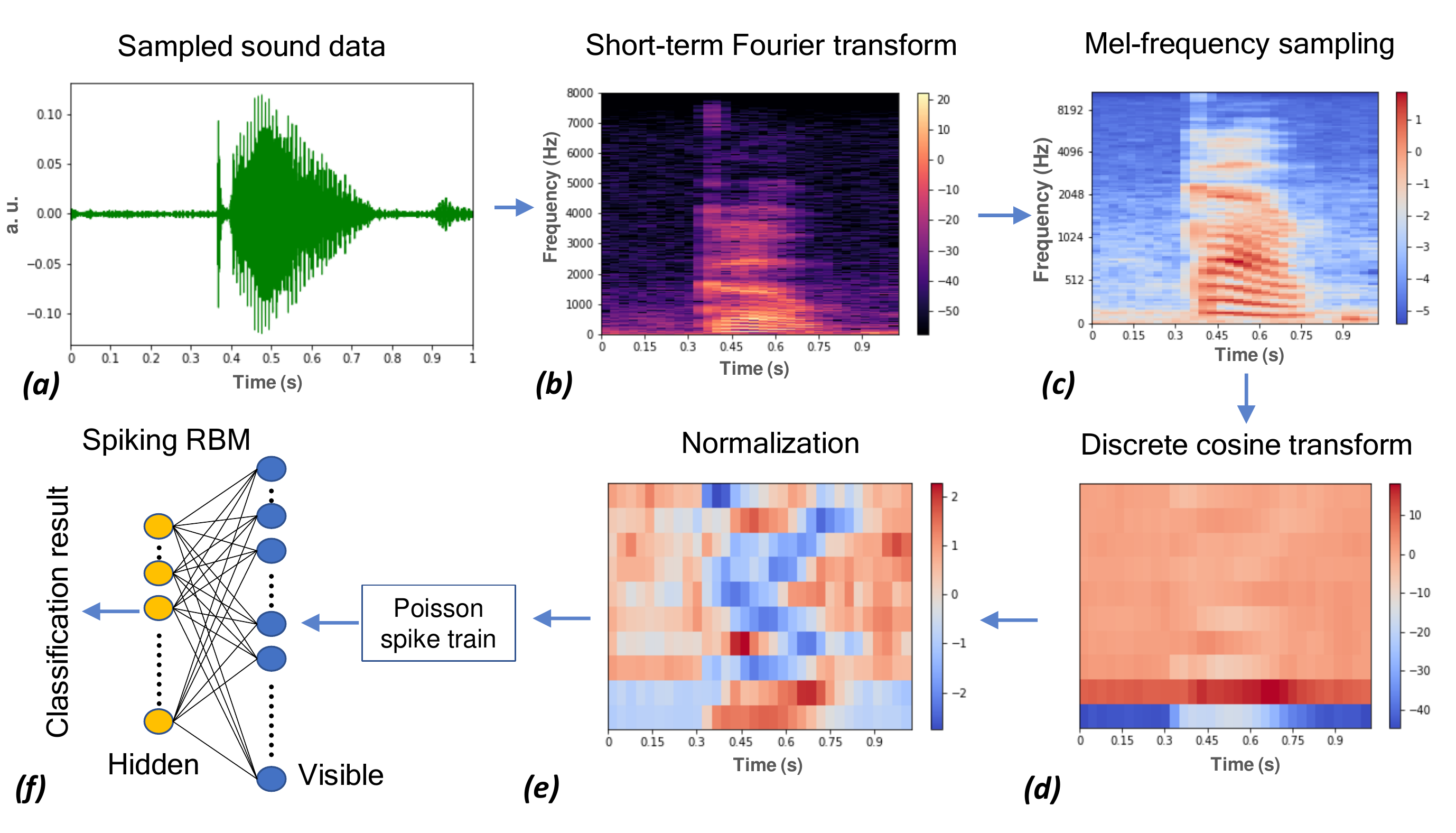}
\caption{The step-by-step procedure followed to demonstrate speech recognition. (a) the raw sound data is first converted into (b) a spectrogram by applying short-term Fourier transforms, (c) and then into a Mel-frequency sampled image. (d) The image is then compressed using the discrete cosine transforms followed by (e) normalization problem. The final normalized output is provided as input to the (f) spiking RBMs.}
\label{fig:Fig_4}
\addtocounter{figure}{0}
\end{figure*}

Currently, in most of the automatic speech recognition systems, the sound data is first converted into MFCC images \cite{Mukh2020}, and the images are fed as inputs to the neural networks. Specifically, four main steps listed below are involved in generating the MFCC images.
\begin{enumerate}
\item Generate a spectrogram (Fig.~\ref{fig:Fig_4}~(b)) for the sound data of an audio file (Fig.~\ref{fig:Fig_4}~(a)) using Short-Time-Fourier Transforms (STFT). The time-varying sound waves are divided into several small overlapping time frames. The frequencies of sound waves in each time frame are then calculated using the fast Fourier transforms. Note that depending on the size of the time frame and the extent of overlap between the two adjacent frames, the spectrogram can either have better time resolution or frequency resolution.  
\item Perform the Mel-frequency sampling on the output spectrogram (Fig.~\ref{fig:Fig_4}~(b)). This sampling re-scales the frequency axis of the spectrogram and emphasizes more on the frequency information in the human's hearing range. 

\item Compress the Mel-sampling output (Fig.~\ref{fig:Fig_4}~(b)) using discrete cosine transforms. This step removes redundant information in the Mel-sampling output.  

\item Finally, normalize the compressed output (Fig.~\ref{fig:Fig_4}~(c)). This step reduces the influence of background noise and cancels out the differences in feature maps between different speakers. 
\end{enumerate}

As shown in Fig.~\ref{fig:Fig_4}~(e), the final output of these steps is fed as input to the spiking RBMs \cite{smolensky}. RBMs are bi-layer stochastic neural networks with neurons in one layer connected to all the neurons in the other layer, but not connected to neurons within the layer. As shown in Fig.~\ref{fig:Fig_5}, neurons in the first (i.e. visible) layer are divided into three categories-image, label, and bias neurons. The size of input images determines the number of image neurons required in this layer. Similarly, the total number of labels under classification determines the number of label neurons needed. On the other hand, neurons in the second (i.e. hidden) layer are divided into two categories-hidden neurons and bias neurons (See Fig.~\ref{fig:Fig_5}). The hidden neurons learn the features of input images and the number of hidden neurons required depends on the total number of weights required to achieve higher accuracies. Finally, the number of bias neurons required (i.e. in both the visible and hidden layers) needs to be tuned to achieve higher accuracies. 
\begin{figure*}[h]
\centering
\includegraphics[width=5.5in]{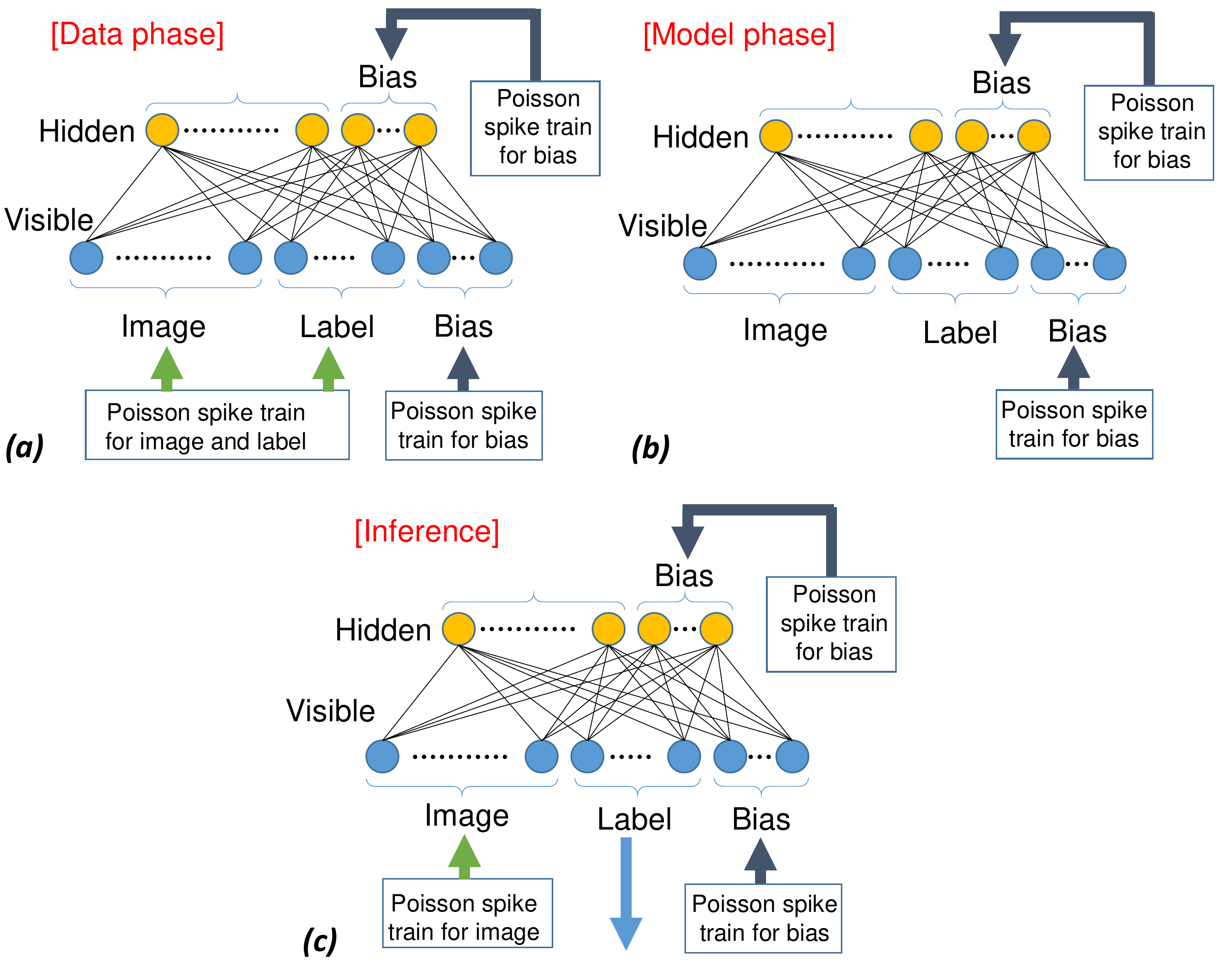}
\caption{Network structure of spiking RBMs with image, label, hidden, and bias neurons. (a) During the data phase in training, the Poisson spike trains are provided as inputs to image, label, and bias neurons. (b) During the model phase in training, only bias neurons receive the Poisson spike trains. (c) During the inference process, an image is provided as input to the image neurons (i.e. in form of the Poisson spike trains) and the classification output is obtained by counting the spikes fired by label neurons.}
\label{fig:Fig_5}
\addtocounter{figure}{0}
\end{figure*}

The spiking RBMs are trained with the STDP-based event-driven CD algorithm \cite{Emre2014}. Specifically, training is performed in two phases-data phase (Fig.~\ref{fig:Fig_5}~(a)) and model phase (Fig.~\ref{fig:Fig_5}~(b)). In the data phase, images and their labels will be given as inputs to the visible neurons in the form of Poisson spike trains. High (low) pixel values in image results in high (low) spiking rates of corresponding image neurons. Further, if an image belongs to a particular class, only the label neurons related to that class will have high spiking rates and all the others will have low spiking rates. All the bias neurons in visible and hidden layers receive Poisson spike trains with high spiking rates. In the data phase, these externally generated spike trains will propagate from visible to the hidden layer and the weights will be updated positively. Next, in the model phase, no external input spikes are provided to the visible/hidden layer neurons except for the bias neurons. Only the internal spikes and the bias neuron spikes will propagate between the two layers and the weights will be updated negatively. When learning converges, there won’t be any further change in the weight be it in the data phase or model phase. 

On the other hand, to perform the inference (Fig.~\ref{fig:Fig_5}~(c)), Poisson spike trains of input images will be provided to the visible neurons. The spikes fired by all the label neurons during the inference period will be counted. The label neurons firing more number of spikes compared to others would be considered as the classification output. For example, when an MFCC image of speech command-``stop'' is given as input and the ``stop'' label neurons fired more spikes than others, the classification output will be ``stop''. In contrast, if the ``go'' label neurons fire more spikes than others, the classification output will be ``go''.
\begin{figure}[h]
\centering
\includegraphics[width=3.5in]{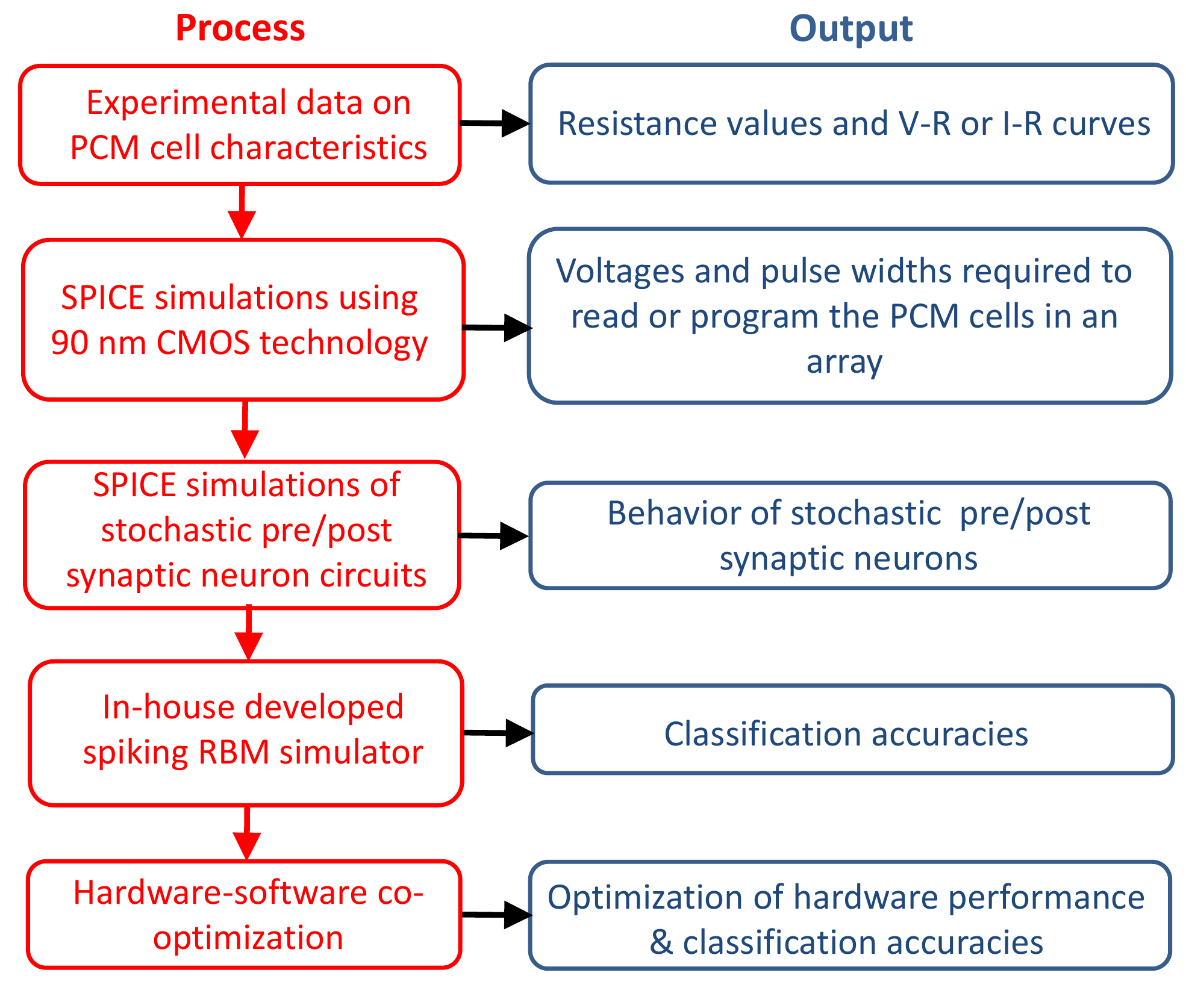}
\caption{The step-by-step procedure followed in the hardware-aware spiking RBM simulator (i.e. indicated by the red-colored text). The output of each process in the flowchart is provided (i.e. indicated by blue-colored text).}
\label{fig:Fig_6}
\addtocounter{figure}{0}
\end{figure}

Note that all the operations required to be performed in spiking RBMs can be accelerated using the Raven (Fig.~\ref{fig:Fig_3}) introduced in Section.~\ref{sec:sec_III}. The presynaptic neurons connected to the left-side of the synaptic array can be used as visible neurons, whereas the postsynaptic neurons connected at the bottom of the synaptic array can be used as hidden neurons. The input (output) spike trains can be provided (read) using the external spike input (output) circuitry. During the data phase of training, the pulse timings of $STDP\_WL\_R_{p}$ and $STDP\_WL\_R_{n}$ can be configured as shown in Fig.~\ref{fig:Fig_2_c}~(b). In such configuration, $R_{p}$ will decrease, and $R_{n}$ will increase and the weight will be updated positively. Similarly, during the model phase of training, the pulse timings of $STDP\_WL\_R_{p}$ and $STDP\_WL\_R_{n}$ can be configured as shown in Fig.~\ref{fig:Fig_2_c}~(c). In such a configuration, $R_{p}$ will increase, and $R_{n}$ will decrease and the weight will be updated negatively. Finally, during the inference, the lines, $STDP\_BL$, $STDP\_WL\_R_{p}$, and $STDP\_WL\_R_{n}$ can be disabled and no spikes will propagate through them. Input spike trains of images will be provided to the visible neurons and concurrently, the spikes fired by all the label neurons will be counted using external counters.
\begin{figure*}[hb!]
\centering
\includegraphics[width=5in]{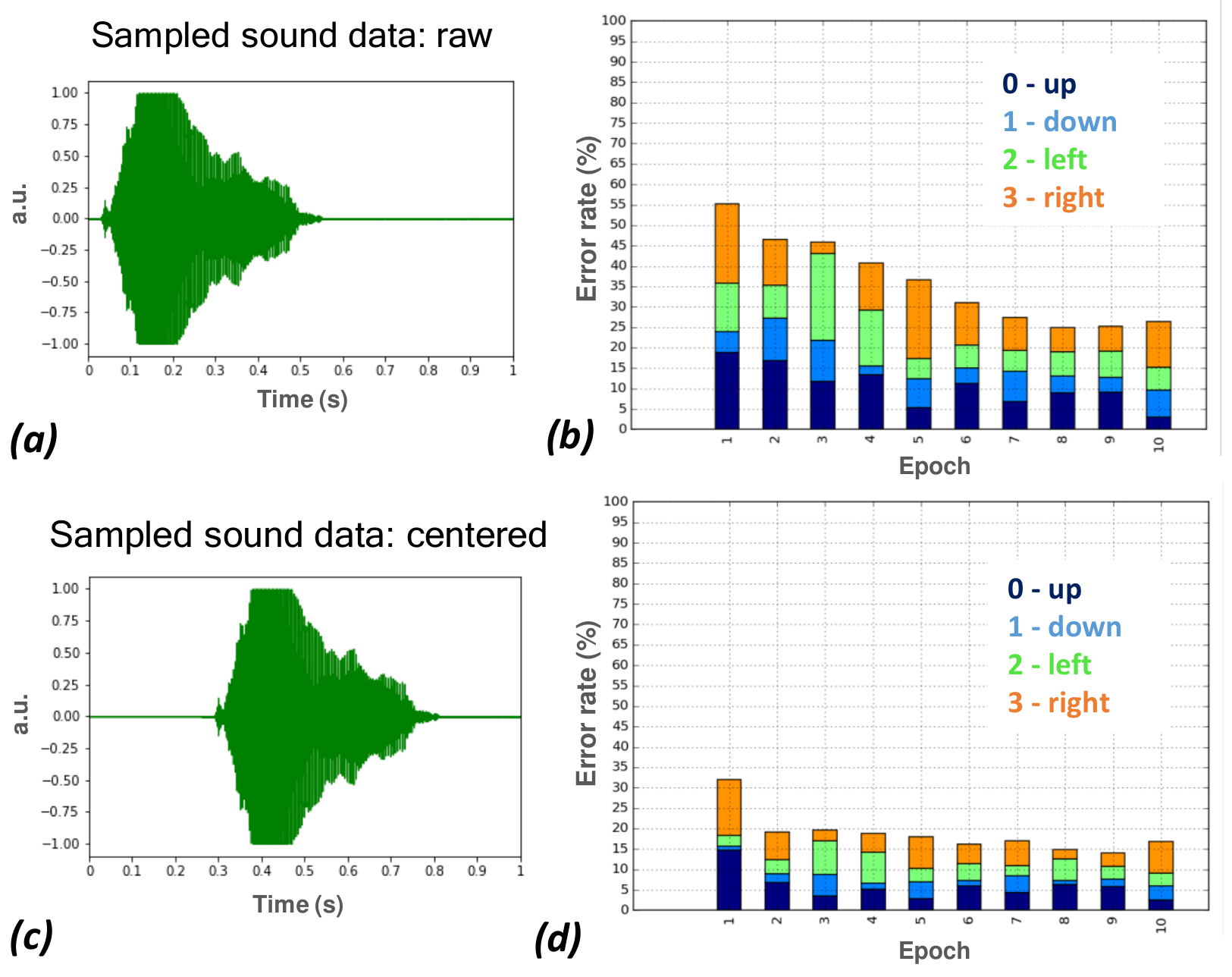}
\caption{(a) sample sound data with the utterance of command in first 0.5 s and (b) the test error rates of four speech commands when the raw sound data is used for training and inference. In each epoch, the average error rate of four speech commands is denoted by the height of a vertical bar. The four colors in a bar represent the contribution of four commands to the average error rate. (c) the modified sound data with utterance centered at 0.5~s and (d) the test error rates of four speech commands when the modified sound data is used for training and inference.}
\label{fig:Fig_7}
\addtocounter{figure}{0}
\end{figure*}

\section{PCM Hardware-aware Spiking RBM Simulator}
\label{sec:sec_IV}

We will now discuss the PCM hardware-aware spiking RBM simulator developed to demonstrate the speech command recognition in this work. 

Based on the earlier works on event-driven CD in spiking neuromorphic systems \cite{Emre2014}, we first developed a spiking RBM simulator that can take sound data as input and perform training and inference operations on the data. This simulator needs several input parameters such as spiking rate, magnitude and pulse width of spikes, equilibrium/rest potential, alpha parameter to update the potential, threshold potential, refractory time, and leak time constant. To estimate these parameters and to take the hardware characteristics/limitations into account, we followed the step-by-step procedure shown in Fig.~\ref{fig:Fig_6}. First, the characteristics of PCM cells such as minimum and maximum resistance values, and current/voltage versus resistance curves are extracted from the experimental data. Next, depending on the size of the synaptic array, the voltages and pulse widths required to read and program the synaptic weights are estimated using SPICE circuit simulations. The behavior of pre and post-synaptic neuron circuits is then studied using SPICE simulations and the abovementioned parameters are estimated and provided as inputs to the spiking RBM simulator, which provides the classification accuracies. Finally, the hardware-software co-optimization is performed based on the classification accuracies and the performance evaluation of circuits in SPICE.
\begin{table}[hbtp]
\caption{Spiking RBM Parameters}
\begin{center}
\begin{tabular}{|c|c|}
\hline
\textbf{Quantity}&\multicolumn{1}{|c|}{\textbf{Value}} \\
\hline
Visible neurons & 384 \\
Label neurons in visible layer & 20 \\
Bias neurons in visible layer & 8 \\
Hidden neurons in hidden layer & 500 \\
Bias neurons in hidden layer & 8 \\
Reset/equilibrium potential value & 0\\
Alpha parameter for updating the potential & 0.06\\
Threshold potential value & 1\\
Leak time constant & 1 ms \\
Refractory time period & 4 ms \\
Spiking rate & 20 Hz \\
\hline
\end{tabular}
\label{tab:tab_1}
\end{center}
\end{table}
\begin{table}[hbtp]
\caption{Voltages and pulse widths used in the proposed hardware}
\begin{center}
\begin{tabular}{|c|c|}
\hline
\textbf{Quantity}&\multicolumn{1}{|c|}{\textbf{Value}} \\
\hline
VSBL & 2.5~V \\
VBWL, VLWL & 1.2~V \\
VWLL & 0.75~V \\
VWLH & 1.4~V \\
a & 9.1 $\mu$s (min) - 2040 $\mu$s (max) \\
b & 4.4 $\mu$s (min) - 85.8 $\mu$s (max) \\
c & 20 ns (min) - 79 ns (max) \\
d & 27 $\mu$s (min) - 4251.6 $\mu$s (max)\\
\hline
\end{tabular}
\label{tab:tab_new}
\end{center}
\end{table}

\section{Results and Discussion}
\label{sec:sec_V}

Using the simulator introduced in Section~\ref{sec:sec_IV}, we will now study the feasibility of performing on-chip training and inference using Raven.  

First, the audio files are converted into MFCC images by choosing the size of each time frame in STFT as 160 ms and the overlap between two adjacent frames as 120 ms. Also, 22 frequency bins are used in DCTs. As a result, the output MFCC images have a size of 22$\times$22 pixels. Next, these MFCC images are provided as inputs to the spiking RBMs simulator with network parameters tabulated in Table~\ref{tab:tab_1} and the magnitude and pulse widths of spikes used are tabulated in Table~\ref{tab:tab_new}. When a set of four speech commands-up, down, left, and right is considered for training and inference, the best test error rate is found to be 25\%. Fig.~\ref{fig:Fig_7}~(b) shows the test error rates observed in each epoch.

The high error rates (See Fig.~\ref{fig:Fig_7}~(b)) arise due to the differences in the exact time at which the commands are uttered in a second. For example, in the sample sound data shown in Fig.~\ref{fig:Fig_7}~(a), the utterance occurred in the first half of a second. If such sound data is used to create an MFCC image, the extracted features will be on the left side of the image (not shown here). Similarly, if the sound data is present in the second half of a second, the extracted features will be on the right side of the image. We found that such variations in the position of features lead to high classification error rates. To resolve this problem, we modified the timing data of each audio file in such a way that the utterance always occur around 0.5~s (as shown in Fig.~\ref{fig:Fig_7}~(c)). As a result, the best test error rate got reduced to 15\% as shown in Fig.~\ref{fig:Fig_7}~(d). 
\begin{figure}[h]
\centering
\includegraphics[width=3.75in]{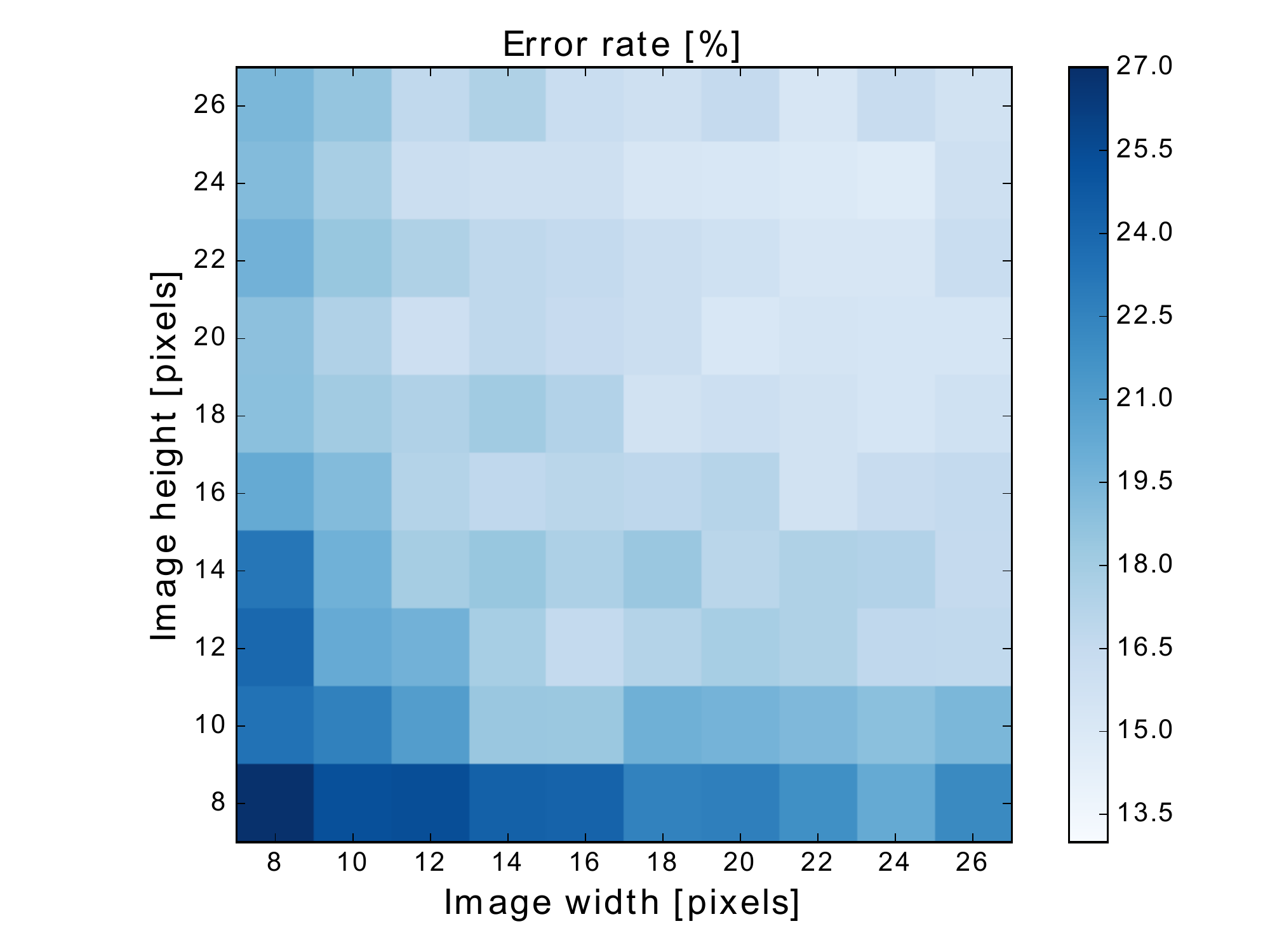}
\caption{Dependence of test error rates on the image width and height. Color bar represents the error rate percentages.}
\label{fig:Fig_8}
\addtocounter{figure}{0}
\end{figure}

Moreover, as discussed in Section~\ref{sec:sec_III}, the size of input images determine the number of image neurons required and thereby the total number of weights in the network. Furthermore, depending on the image size, the MFCC image can either have better time resolution or frequency resolution, but not both. Therefore, it is crucial to find out the optimum image size required to achieve low classification error rates. The phase diagram is shown in Fig.~\ref{fig:Fig_8} depicts the dependence of error rates on the input image size. As shown in Fig.~\ref{fig:Fig_8}, low error rates are obtained when the image width/height is between 20 to 24 pixels. 

To further reduce the error rates, it is proven in the literature that multiple images of different sizes can be placed side by side and provided as input to the neural networks \cite{doroteo2018}. We also studied this possibility and found that an error rate of 13.5\% can be achieved by using two images of size-16$\times$16 and 8$\times$16. Next, using such a two image configuration, we estimated the classification accuracies for different sets of speech commands and compared them with the results obtained from the state-of-the-art convolution neural networks (CNNs) \cite{warden2018speech, li2019speech, patil2019}. As shown in Table~\ref{tab:tab_2}, the minimum (maximum) accuracy difference between CNNs and our work is found to be 5.12\% (18.93\%). Such an accuracy difference is expected as SNNs with STDP-based learning generally have moderate classification performance when compared to DNNs trained with backpropagation. Currently, there are several on-going research efforts to close the accuracy gap between DNNs and SNNs \cite{KIM2020107742,DENG2020294,falez2020improving}.
\begin{table}[hb!]
\caption{Comparison of classification accuracies between our work and the state-of-the-art CNNs.}
\begin{center}
\begin{tabular}{|c|c|c|}
\hline
\textbf{Speech commands}&\multicolumn{1}{|c|}{\textbf{CNNs}}&\multicolumn{1}{|c|}{\textbf{our work}} \\
\hline
up, down, left, right & 94.5\% & 86.4\% \\
\hline
bed, cat, happy, bird, five & 91.36\% & 86.24\% \\
\hline
spoken digits [0-9] & 90\% & 78.36\% \\
\hline
stop, go, left, right, on, & & \\
off, up, down, yes, no, & 88.2\% & 69.27\%\\
silence, unknown & & \\
\hline
\end{tabular}
\label{tab:tab_2}
\end{center}
\end{table}
\begin{table}[htbp!]
\caption{Description of each layer in the FCNN considered in this work.}
\begin{center}
\begin{tabular}{|c|c|c|}
\hline
\textbf{Description}&\multicolumn{1}{|c|}{\textbf{W$\times$H$\times$IFs}}&\multicolumn{1}{|c|}{\textbf{W$\times$H$\times$OFs}} \\
\hline
Conv1 & 24$\times$16$\times$1 & 22$\times$14$\times$64 \\
\hline
Maxpool1 & 22$\times$14$\times$64 & 11$\times$7$\times$64 \\
\hline
Conv2 & 11$\times$7$\times$64 & 9$\times$5$\times$52 \\
\hline
Maxpool2 & 9$\times$5$\times$52 & 4$\times$2$\times$52 \\
\hline
Conv3 & 4$\times$2$\times$52 & 3$\times$1$\times$36 \\
\hline
Maxpool3 & 3$\times$1$\times$36 & 1$\times$1$\times$36 \\
\hline
Conv4 & 1$\times$1$\times$36 & 1$\times$1$\times$10 \\
\hline
Dense & 1$\times$1$\times$10 & 10 \\
\hline
\end{tabular}
\label{tab:tab_3}
\end{center}
\end{table}

Next, to compare the memory and computational requirements of our work with the CNNs at iso-accuracies, we implemented a fully-convolutional neural network (FCNN) \cite{long2015} with 8 layers as shown in Table~\ref{tab:tab_3}. In Table~\ref{tab:tab_3}, $\it{W}$ and $\it{H}$ represent the width and height of the feature maps, $\it{IFs}$ represent the number of input feature maps provided to each layer, $\it{OFs}$ represent the number of output feature maps extracted from each layer. We optimized and the parameters tabulated in Table~\ref{tab:tab_3} to achieve the same classification accuracies as our work. The backpropagation algorithm with stochastic gradient descent is used to train the FCNNs \cite{rumelhart1986}, while the spiking RBMs are trained using event-driven CD algorithm. Note that unlike the spiking RBMs, weights in FCNNs are trained using 32-bit floating-point numbers. The number of parameters, spikes/multiply-and-accumulate operations (MACs), and epochs required to obtain iso-accuracies are estimated and tabulated in Table~\ref{tab:tab_4}. MACs are the fundamental operations required by the CNNs. As shown in Table~\ref{tab:tab_4}, the number of MACs performed in the FCNNs during training and inference is 269.23$\times$ and 70.36$\times$ greater than the number of spikes generated in the spiking RBMs, respectively. Due to such low computational requirements, the spiking RBM implementation can be more suitable for edge applications, in which accuracies may not be of paramount importance. 
\begin{table}[ht!]
\caption{Iso-accuracy comparison of spiking RBMs and FCNN network requirements.}
\begin{center}
\begin{tabular}{|c|c|c|}
\hline
\textbf{Quantity}&\multicolumn{1}{|c|}{\textbf{Our work}}&\multicolumn{1}{|c|}{\textbf{FCNNs}} \\
\hline
Training method & Event-driven CD & Backpropagation \\
 &  &
with SGD \\
\hline
Total number of training
images & 5000 & 5000 \\
\hline
Size of input image & 24$\times$16 & 24$\times$16 \\
\hline
Parameters & 209296 & 38476 \\
\hline
Epoch & 6 (batch size = 1) & 3 (batch size = 1) \\
\hline
Test accuracy & 78.36\% & 78.96\% \\
\hline
Spikes/MACs during inference & 0.022 M & 1.548 M \\
\hline
Spikes/MACs during training & 172.51 M & 46.445 B \\
\hline
\end{tabular}
\label{tab:tab_4}
\end{center}
\end{table}

Finally, using the SPICE simulations, we estimated the power and latencies consumed by the Raven circuits and architectures during the training and inference operations. The power and latencies consumed during the training of 5000 MFCC images are estimated to be 30~$\mu$W (7~$\mu$W of active power and 23~$\mu$W of static power) and 3000 sec, respectively. Also, the power and latency consumed for an inference operation on Raven are estimated to be 28~$\mu$W (5~$\mu$W of active power and 23~$\mu$W of static power) and 0.45 sec, respectively. Note that we used the 90~nm CMOS technology for this work. 

\section{Conclusion}
\label{sec:sec_VI}

In summary, the ultra-low-power on-chip training and inference of speech commands are demonstrated using the phase change memory (PCM)-based synaptic arrays. The power and latencies consumed during on-chip training (inference) are estimated to be 30~$\mu$W and 3000 sec (28~$\mu$W and 0.45 sec). Furthermore, at iso-accuracies, the number of multiply-and-accumulate operations (MACs) needed during the training of a deep neural network (DNN) model is found to be 269.23$\times$ greater than the number of spikes required in our work. Similarly, during inference, the number of MACs needed during the inference of DNN is 70.36$\times$ greater than the number of spikes required in our work. Overall, due to such low power and computational requirements, the PCM-based synaptic arrays can be promising candidates for enabling AI at the edge.

% use section* for acknowledgment
\section*{Acknowledgment}

The authors would like to express special thanks to Seiji Munetoh, Atsuya Okazaki, and Akiyo Nomura for their valuable and insightful comments.

% Can use something like this to put references on a page
% by themselves when using endfloat and the captionsoff option.
\ifCLASSOPTIONcaptionsoff
  \newpage
\fi

% trigger a \newpage just before the given reference
% number - used to balance the columns on the last page
% adjust value as needed - may need to be readjusted if
% the document is modified later
%\IEEEtriggeratref{8}
% The "triggered" command can be changed if desired:
%\IEEEtriggercmd{\enlargethispage{-5in}}

% references section

% can use a bibliography generated by BibTeX as a .bbl file
% BibTeX documentation can be easily obtained at:
% http://mirror.ctan.org/biblio/bibtex/contrib/doc/
% The IEEEtran BibTeX style support page is at:
% http://www.michaelshell.org/tex/ieeetran/bibtex/
%\bibliographystyle{IEEEtran}
% argument is your BibTeX string definitions and bibliography database(s)
%\bibliography{IEEEabrv,../bib/paper}
%
% <OR> manually copy in the resultant .bbl file
% set second argument of \begin to the number of references
% (used to reserve space for the reference number labels box)

\bibliographystyle{IEEEtran}
\bibliography{IEEEabrv,main.bib}

% biography section
% 
% If you have an EPS/PDF photo (graphicx package needed) extra braces are
% needed around the contents of the optional argument to biography to prevent
% the LaTeX parser from getting confused when it sees the complicated
% \includegraphics command within an optional argument. (You could create
% your own custom macro containing the \includegraphics command to make things
% simpler here.)
%\begin{IEEEbiography}[{\includegraphics[width=1in,height=1.25in,clip,keepaspectratio]{mshell}}]{Michael Shell}
% or if you just want to reserve a space for a photo:

% insert where needed to balance the two columns on the last page with
% biographies
%\newpage

% You can push biographies down or up by placing
% a \vfill before or after them. The appropriate
% use of \vfill depends on what kind of text is
% on the last page and whether or not the columns
% are being equalized.

%\vfill

% Can be used to pull up biographies so that the bottom of the last one
% is flush with the other column.
%\enlargethispage{-5in}

% that's all folks
\end{document}